\documentclass[runningheads,a4paper]{llncs}

\usepackage{amssymb,amsmath}
\usepackage{graphicx}

\usepackage{url}
\urldef{\mailsa}\path{naldi@disp.uniroma2.it}
\urldef{\mailsb}\path|dacquisto@ing.uniroma2.it|
\newcommand{\keywords}[1]{\par\addvspace\baselineskip
\noindent\keywordname\enspace\ignorespaces#1}

\begin{document}

\mainmatter  

\title{Differential Privacy: An Estimation Theory-Based  Method for Choosing Epsilon}

\titlerunning{Laplace noise for differential privacy}

%
%
\author{Maurizio Naldi%
\and Giuseppe D'Acquisto}
\authorrunning{M. Naldi - G. D'Acquisto}

\institute{Universit\`{a} di Roma Tor Vergata\\Department of Computer Science and Civil Engineering\\
Via del Politecnico 1, 00133 Roma, Italy\\
\mailsa\\
\mailsb\\
}

%
%

\toctitle{Lecture Notes in Computer Science}
\tocauthor{Authors' Instructions}
\maketitle

\begin{abstract}
Differential privacy is achieved by the introduction of Laplacian noise in the response to a query, establishing a precise trade-off between the level of differential privacy and the accuracy of the database response (via the amount of noise introduced). However, the amount of noise to add is typically defined through the scale parameter of the Laplace distribution, whose use may not be so intuitive. In this paper we propose to use two parameters instead, related to the notion of interval estimation, which provide a more intuitive picture of how precisely the true output of a counting query may be gauged from the noise-polluted one (hence, how much the individual's privacy is protected).
\keywords{Statistical databases; Differential privacy; Anonymization.}
\end{abstract}

\section{Introduction}
In statistical databases, records may include personal details, but responses are provided for queries concerning aggregate data, i.e., as statistics. Though the access to individual records may be denied, it is possible to use a combination of aggregate queries to obtain information about a single individual. In order to guarantee the privacy of users whose data are contained in statistical databases, the notion of differential privacy has been introduced \cite{dwork2011}.

Through differential privacy, the probability of obtaining the same response to two queries where the actual values differ instead by 1 (e.g., differ for 1 individual) is lower bounded by $e^{-\epsilon}$, where $\epsilon$ is the differential privacy level, and the mechanism is said to be $\epsilon$-differentially private. By setting $\epsilon$ very close to 0, the responses to two queries as described above can be made practically indistinguishable from each other, so that their combination cannot be used to infer the datum concerning that individual. However, setting $\epsilon$ too close to 0 may make the queries useless, since the response could provide no clue at all about the actual values contained in the database (see \cite{naldi2014differential} for the possibility of using Bayes theorem to refine the estimation of the true value).

Setting the correct level of differential privacy is therefore essential for the actual use of the mechanism, allowing to strike a balance between the wish for obfuscation and the need for statistical information. Unfortunately, there are no consistent indications in the literature as to the value (or the range of values) that should be used for $\epsilon$. In \cite{dwork2011} it is stated that its selection is a social issue, and very different values are employed, such as 0.01, 0.1, or, in some cases, $\ln$2 or $\ln$3 (spanning more than two orders of magnitude). Values ranging from 0.05 to 0.2 are used in \cite{Sarwate2012}.

In \cite{hsu2014} an economic method has been proposed to set the right value for $\epsilon$, where two parties have conflicting views about the use of the data: the \textit{data analyst} wishes to conduct a study with an accuracy as high as possible, while an \textit{individual} contributes its information to the database in return for a compensation. Since it is the analyst's duty to compensate the individual for the data, the right value for $\epsilon$ is the result of a balance between the wish for accuracy and the budget available to compensate the individual. This mechanism relies however on the adoption of a specific use model for the data (the analyst vs individual, described above), which may not be the only one. For example, in \cite{MNPrinf15} and \cite{MNcns15} a different use of counting queries is envisaged in dual supply markets, where the individuals contributing their data are actually privacy-aware sellers that do not wish to divulge the presence and amount of a particular item they have in stock

In this paper, we propose a different method to determine the right value for $\epsilon$. This method provides a definition of the level of differential privacy in terms of the accuracy that may be obtained in an interval estimation of the true value, and is embodied by two parameters, namely the confidence interval and the confidence level, that are more intuitive than the original $\epsilon$.

The paper is organized as follows. In Section \ref{noise}, we review the notion of differential privacy and its implementation through the addition of Laplace noise. In Section \ref{interval}, we relate the level of differential privacy to the two parameters we have introduced above and provide the way to obtain the desired value of $\epsilon$.

\section{Laplace noise and differential privacy}
\label{noise}
In this section, we deal with the notion of differential privacy and see how the use of a Laplace-distributed noise is related to it.

In \cite{dwork2011} it is stated that "achieving differential privacy revolves around hiding the presence or absence of a single individual." In the same paper it is shown that differential privacy can be achieved for a counting query by adding Laplace-distributed noise to the query output. Namely, if noise following a Laplace density function
\begin{equation}
f_{N}(x) = \frac{\lambda}{2}e^{-\lambda\vert x \vert}
\end{equation}
is added, we obtain $\lambda$-differential privacy (in papers on differential privacy the symbol $\epsilon$ is typically used with the same meaning as $\lambda$; here we stick to $\lambda$ as typical in the terminology concerning the Laplace distribution). In fact, the definition of differential privacy requires the added random noise to be such that for any two values $w$ and $z$ differing by 1 ($\vert w - z \vert =1$) we have $f_{N}(w) \le e^{\epsilon}f_{N}(z)$. For dealing at the same time with the cases where $w<z$ or $w>z$, that inequality may be rewritten as
\begin{equation}
\label{diffprivdef}
e^{-\epsilon} \le \frac{f_{N}(w)}{f_{N}(z)}  \le e^{\epsilon},
\end{equation}
so that the noise ie capable of hiding the difference between values that the query output may take on a pair of databases that differ in only one row. For example, let's consider the  two cases where: a) the true value is 10 and the added noise is 2; b) the true value is 11 and the added noise is 1. In both cases we end up with an output equal to 12. If we ask ourselves what the probability is of getting that response when the true value is either 10 or 11, we must consider the probability densities of the random noise for the values 2 and 1 respectively. If we set $\epsilon = 0.01$ and indicate the true value by $c$ and the noisy response by $\hat{c}$, we have 
\begin{equation}
\begin{split}
e^{-0.01} &\le \frac{\mathbb{P}[\hat{c}=12 \vert c=10]}{\mathbb{P}[\hat{c}=12 \vert c=11]}  \le e^{0.01}\\
0.99 &\le \frac{\mathbb{P}[\hat{c}=12 \vert c=10]}{\mathbb{P}[\hat{c}=12 \vert c=11]}  \le 1.01 ,
\end{split}
\end{equation}
i.e. we cannot practically distinguish if the true value was 10 or 11. The choice of the Laplace distribution with $\lambda = \epsilon$ satisfies the inequality (\ref{diffprivdef}) and therefore provides differential privacy. In fact, for $w>0$ (but the same expression can be obtained for $w<0$) we have
\begin{equation}
\frac{f_{N}(w)}{f_{N}(w+1)} = \frac{\frac{\epsilon}{2}e^{-\epsilon\vert w \vert}}{\frac{\epsilon}{2}e^{-\epsilon\vert w+1 \vert}} = e^{\epsilon}  
\end{equation}

It is to be noted that the level of privacy protection is higher the lower $\epsilon$, since that guarantees that both bounds in the inequality (\ref{diffprivdef}) are very close to 1.

\section{Interval estimation of true value}
\label{interval}
In Section \ref{noise} we have seen how differential privacy is achieved through the addition of noise following a Laplace distribution, so that the level of privacy is embodied by the single parameter $\lambda$, i.e., the scale parameter of the Laplace distribution. However, the use of that parameter to define the desired level of privacy is not very intuitive. In this section we describe how it can be related to a different way of describing how effective the obfuscation level is in protecting the individual privacy.

Since the true value $c$ provided by a counting query is masked through the addition of Laplace noise, and the value output by the differential privacy mechanism is 
\begin{equation}
\hat{c} = c+ N,
\end{equation}
where $N$ is the Laplace noise, that mechanism can be reversed to obtain an estimate $X$ of the true value. We have at our disposal the single value $\hat{c}$ output by the differential privacy mechanism, so that the point estimator would be
\begin{equation}
\label{pointest}
X = \hat{c}.
\end{equation}

The point estimate obtained by Equation (\ref{pointest}) is of little use to evaluate how far the declared value may be from the true one (and therefore to assess how effective the differential privacy mechanism is). We can instead turn to interval estimation, which, rather than providing a single value, gives us a range of a values that is likely to contain the true value, replacing a precise statement whose probability is 0 with an approximate statement with positive probability. The estimate associates a probability figure to that range, so the level of obfuscation introduced is more clearly understood. Since the Laplace probability density function is symmetrical, and we have no reason to discriminate between the two directions in which we can err, we aim at a description of the privacy protection level through a mathematical statement of the following form
\begin{equation}
\label{intest}
\mathbb{P}[\hat{c}-wc < c <\hat{c}+wc] = p,
\end{equation}
where the probability value is typically chosen so high as to have a high confidence that the true value is within $\pm w\%$ of the declared one. The description is now summarized  by the two parameters $w$ and $p$, which are a measure respectively of the confidence interval width and the confidence level. For a fixed $p$, the larger $w$ the higher the level of protection privacy, since the recipient of the query output finds it more difficult to estimate the true value.

Equation (\ref{intest}) can be reversed to obtain a condition on the noise probability parameters:
\begin{equation}
\label{}
\begin{split}
 \mathbb{P}[\hat{c}-wc < c <\hat{c}+wc] &= \mathbb{P}[c-wc < \hat{c} < c+wc] = \mathbb{P}[(1-w)c < c+N < (1+w)c]\\
  &=\mathbb{P}[-wc < N < w+c] = p
 \end{split}
 \end{equation} 

Since the noise follows a Laplace probability distribution function, the confidence level $p$ represented by Equation (\ref{intest}) can be written as
\begin{equation}
\label{privpar}
\begin{split}
p &=  \mathbb{P}[-wc < N < w+c] = F_{N}(wc) - F_{N}(-wc)\\
& = 1-\frac{1}{2}e^{-\lambda wc} -  \frac{1}{2}e^{-\lambda wc} = 1 - e^{-\lambda wc}.
\end{split}
\end{equation}

The privacy level can then be related to the two parameters $w$ and $p$ (describing the uncertainty surrounding the true value) by solving Equation (\ref{privpar}) for $\lambda$ (and therefore $\epsilon$):
\begin{equation}
\label{choice}
\epsilon = \lambda = - \frac{\ln (1-p)}{wc},
\end{equation}
where we see that we must ask for more differential privacy (i.e., a lower $\epsilon$) as the true vaue $c$ grows, i.e., we must add more noise to achieve the same relative uncertainty.

\begin{figure}
\begin{center}
\includegraphics[scale=0.4]{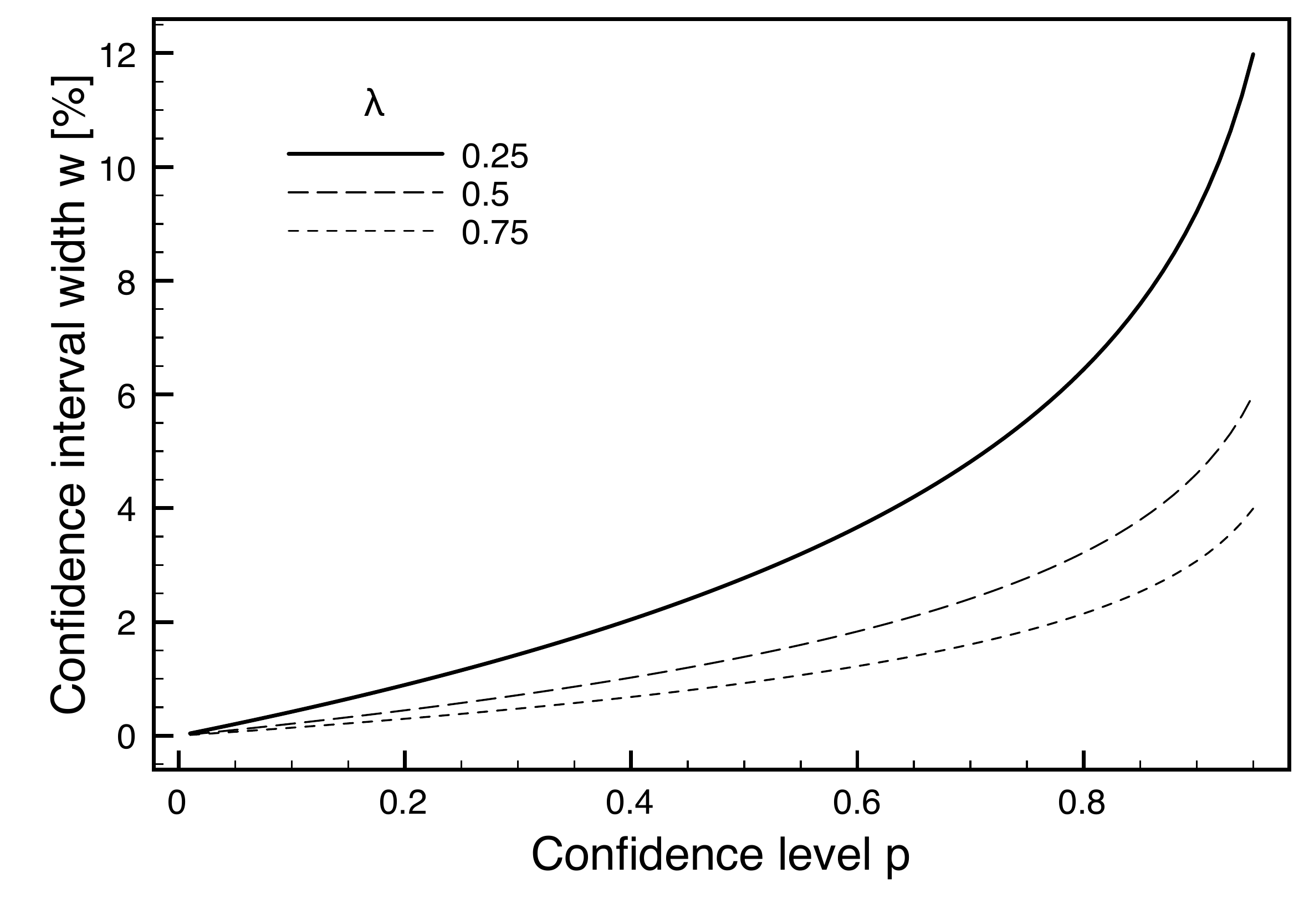} 
\label{fig:isocurve}
\caption{Iso-lambda curves}
\end{center}
\end{figure}

This formula allows us to determine the correct value of privacy level $\lambda$, once we have set the level of uncertainty we wish for the output of the counting query. For example, if the true value is $c=100$ and we wish the query output to be within $\pm 20\%$ of the true value with 80\% probability, the right value is $\lambda = \ln (1-0.8) /(0.2\cdot 100) = 0.08$. Some iso-lambda curves are shown in \figurename~\ref{fig:isocurve} for $c=100$, where it can be seen that lower values of $\lambda$ (granting higher privacy protection) correspond to higher values of the interval relative half-width $w$.

\begin{figure}[h!]
\begin{center}
\includegraphics[scale=0.4]{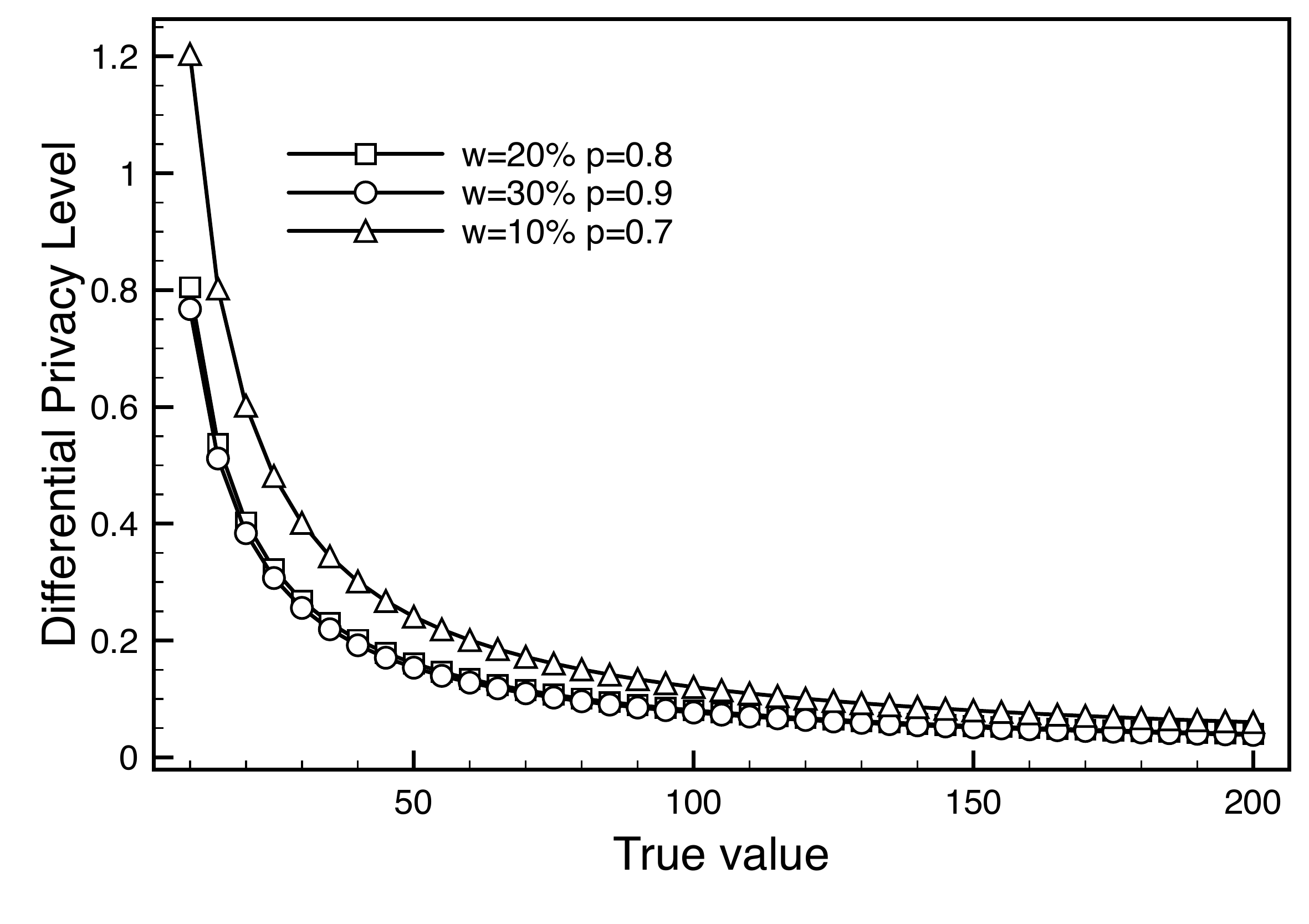} 
\label{fig:invtrue}
\caption{Inverse dependence on the true value}
\end{center}
\end{figure}

The inverse dependence of $\epsilon$ on the true value is shown in \figurename~\ref{fig:invtrue} for three different combinations of confidence interval width and confidence level, where it can be seen how the numerical value of $\epsilon$ must decrease (hence, the amount of noise added must grow) as the true value grows.

\section{Conclusions}
The scale parameter of the Laplace noise distribution, is not such an intuitive means of quantifying the level of differential privacy. The use of two parameters, which describe the probability (confidence level) that the true result of a counting query is within a given range (confidence interval) of the noisy query result, is advocated as providing a more precise picture of the level of differential privacy achieved. The relationship between these two parameters and the Laplace scale parameter is provided.

\bibliographystyle{splncs03}
\bibliography{Bib-privacy}

\end{document}